\documentclass[a4paper,10pt,twocolumn]{article}

\usepackage[english]{babel}
\usepackage[latin1]{inputenc}
\usepackage{graphicx}
\usepackage{rotating}
\usepackage{cite}
\usepackage{upgreek} 
\usepackage{amsfonts}
\usepackage{amssymb}
\usepackage{amsmath}
\usepackage{caption}
\usepackage{authblk}  

\usepackage[backref=page]{hyperref}
\hypersetup{
		colorlinks=true,	      
	  linkcolor=blue,          
    citecolor=blue,        
    urlcolor=green,           
}

\title{Fabry-P\'{e}rot etalon walk-off loss in ring cavities}
\date{\today}
\author[1,2]{Ulrich Eismann}
\affil[1]{Universit\'e Pierre et Marie Curie - Paris VI, 75005 Paris, France}
\affil[2]{now with KEEQuant GmbH, Gebhardtstrasse 28, 90762 F\"urth, Germany}

\begin{document}
\pagestyle{empty}

\maketitle

\renewcommand{\thefootnote}{} 
\footnotetext{Submitted for review in Optics Communications.}

\abstract{
Fabry-P\'erot etalons are widely used for laser frequency control. Inserting them in laser cavities imparts unavoidable walk-off loss and reduces the output power. Here, we treat the technology-relevant case of the walk-off loss of an etalon in a unidirectional ring cavity, which is a standard design of single-frequency lasers. We provide the theory background by discussing the analytic limits. The loss can be efficiently minimized by realignment, or by proper matching of the etalon's surface reflectivity with its refractive index. We numerically calculate the loss in the region uncovered by the analytic limits. We discuss practical laser design considerations, and we perform a tilt-tuning experiment in a single-frequency, solid-state laser setup. 
\bigskip
\vspace*{.5cm}
}

\section{Introduction}
\label{SIntroduction}

Today, single-frequency lasers are the workhorses in numerous technical and scientific applications, such as optical telecommunication, atomic physics, laser spectroscopy, and the emerging quantum technologies. Depending on the type of laser, achieving both stable single-frequency operation and frequency tunability requires specific design ingredients. In the popular case of solid-state lasers, both properties are usually achieved by combining two design features: First, a unidirectional ring laser cavity yields maximum mode competition. 
Second, intracavity frequency selective elements like a Fabry-P\'erot etalon (etalon) significantly increase the cavity round-trip loss for all but the desired 
longitudinal mode\,\cite{Peterson1966}, effectively acting as a frequency filter.

Any laser intracavity optical element will contribute to cavity round-trip loss. In the case of an etalon, this is due to a number of technical imperfections like surface scattering, absorption, or non-parallelism of surfaces. The added loss will reduce the laser output power, or even pull the system below its lasing threshold, and should therefore be minimized. In most practical cases, the technical imperfections can be reduced to a negligible degree by proper design and manufacturing of the components. 

An intrinsic loss process persists, which is caused by lateral walkoff of the laser beam inside the etalon. This loss process cannot be avoided, because the laser cannot be operated without beam walkoff in the etalon, i.e. with the etalon surfaces exactly perpendicular to the intracavity laser beam. In  this case, instead of the etalon acting as a frequency filter only, the laser beam will be reflected back from the etalon into the laser cavity mode, giving rise to a potentially highly unstable multi-coupled-cavity regime of laser operation. Such, a minimum tilt angle has to be maintained to avoid such coupling, which can be estimated to be on the order of the gaussian beam divergence angle
\begin{equation}
\theta_{\rm min} \sim \frac{\lambda}{\pi w_0}
\rm ,
\label{Ethetamin}
\end{equation}
where $\lambda$ is the laser operation wavelength and $w_0$ is the waist radius of the fundamental mode of the laser resonator. 

Furthermore, etalon tilting can be used to obtain tuning of the laser. The frequency tuning is proportional to the square of the tilt angle. 
As we will show here, the walk-off loss shows the same scaling.

For completeness, we want to mention that alternative techniques for enforcing single-frequency operation and tuning of lasers have been demonstrated. Examples are different intracavity frequency filter classes, intracavity second-harmonic generation\,\cite{Eismann2013}, feedback from optics external to the laser cavity\,\cite{Miake2015}, and injection locking \cite{Koch2015}.

The single-pass transmission functions of etalons have been discussed extensively in literature, see for instance \cite{Cotteverte1991,Abu-Safia1994,Nichelatti1995}. Both analytic\,\cite{Danielmeyer1970,Korolev1976, Leeb1975} and numeric calculations\,\cite{Leeb1975} of the loss have been performed for the case of a linear (standing-wave) cavity, where the laser beam passes the etalon twice per cavity round trip.
However, to the best of our knowledge, no calculation of etalon-walk-off insertion loss has been published for ring resonators with a one-way passage through the etalon. 

The analysis presented here can be applied to any resonant ring cavity where an etalon is inserted, for instance in single-frequency optical parametric oscillators\,\cite{Eismann2019}. A similar experimental situation occurs whenever etalons are used as single-pass frequency filters for gaussian beams with subsequent spatial filtering by a pinhole or a single-mode optical fiber. Among applications in spectroscopy and astronomy\,\cite{Murphy2007}, such setups are used in quantum optics to filter single photons of interest from a background which can be many orders of magnitude greater in intensity\,\cite{Tian2016}.

This article is organized as follows: In Section\,\ref{SAnalytic_limits} we derive the analytical expressions in the limits of very large and small walk-off. For the practically important case of small walk-off, we show that the loss is on the same order like the double-pass situation in a linear cavity, and that it can be minimized by realignment of the cavity. We show that by matching of the surface reflectivity and the refractive index, the loss can be minimized even automatically, without manual cavity realignment.
In Section\,\ref{SNumericCalculation}, we compare the analytic results to a numeric optimization covering the full range of walk-off between the analytic limits, and find very good agreement in both limits. 
In Section \,\ref{SPracticalDesignConsiderations} we discuss important practical considerations for low-loss etalon designs. In Section \,\ref{SExperiments} we use a laser setup to perform an etalon tilt-tuning experiment, and find excellent agreement between our theory and the achieved laser output power. We conclude in Section \,\ref{SConclusion}.

\section{Analytic limits}
\label{SAnalytic_limits}

\begin{figure}[]
\centering\includegraphics[width=\columnwidth]{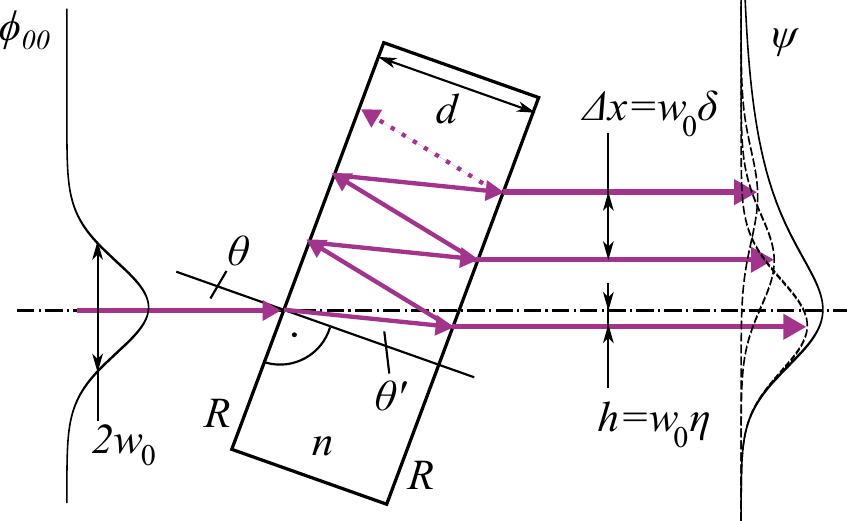}
\caption{(Color online) Schematic sketch of the alternation happening to a gaussian beam (radius $w_0$) when a tilted etalon (tilt angel $\theta$, thickness $d$, surface reflectivity $R$, refractive index $n$) is transmitted. The incident beam travels along the optical axis (dash-dotted line). 
The incident wave function $\phi_{00}$ and the transmitted wave function $\psi$ are plotted in solid lines, and the contribution of several orders of transmitted beams are plotted as a dashed lines. The zeroth-order transmitted beam is shifted laterally by a distance $h$.
}
\label{FEtalon_walkoff}
\end{figure}

The experimental situation under study is depicted in Fig.\,\ref{FEtalon_walkoff}. When the etalon's surface normal vector and the direction of propagation of a Gaussian beam form an angle of incidence $\theta$, a lateral walk-off of subsequent orders of the transmitted waves occurs. We define the normalized walk-off between succeeding orders of reflection
\begin{equation}
\delta = \frac{\Delta x}{w_0} = \frac{2d}{w_0}\tan{\theta'}\cos{\theta} \simeq \frac{2d}{n w_0}  \theta
\rm ,
\label{Edelta}
\end{equation}
where $\Delta x$ is the spatial offset between adjacent etalon transmission orders, $d$ is the etalon's length, and Snell's law reads $\sin(\theta)/\sin(\theta')=n$ with $n$ being the refractive index of the etalon. We note that together with the etalon's surface reflectivity $R$, the set of etalon design parameters $\lbrace R, n, d\rbrace$ can be chosen at will. It is the goal of this article to provide an optimum design choice of parameters under the constraints given by the laser design and the necessary frequency selectivity.

In the case of large walk-off $\delta \gtrsim 1$, the multiple orders of the etalon reflection do not overlap significantly, and one finds for the loss of the strongest transmitted beam
\begin{equation}
L_{\rm max} = 1-(1-R)^2 = 2R-R^2
\rm .
\label{ELmax}
\end{equation}

The field of a power-normalized gaussian beam reads
\begin{equation}
\phi_{00}(x,y) = \sqrt{\frac{2}{\pi w_0^2}}\exp \left( -\frac{x^2 + y^2}{w_0^2} \right)
\rm .
\label{Ephi}
\end{equation}
After passing the etalon, in which the beam experiences multiple reflections, the transmitted wave function reads
\begin{equation}
\psi(x,y) = t_1 t_2 \tau \sum_{j=0}^\infty p^j \phi(x+ j{\Delta x},y)
\rm ,
\label{Epsi}
\end{equation}
where the $t_i$ is the amplitude transmission of surface $i$, $\tau$ is the amplitude transmission of the etalon bulk material (and can in principle also account for further loss), and the round-trip parameter reads $p = r_1 r_2 \tau^2 \exp(-{\rm i} \Phi)$ with $r_i$ being the amplitude reflectivity of surface $i$, and $\Phi$ the round-trip phase. In the resonant case we consider, $\Phi = 0~{\rm mod}(2\pi)$. In order to calculate the loss for the lowest order gaussian mode we consider, we define the spatial overlap integral
\begin{equation}
C = \int_{-\infty}^\infty {\rm d}x {\rm d}y ~\phi_{00}^\ast \,\psi
\rm ,
\label{EC}
\end{equation}
yielding the power loss
\begin{equation}
L = 1 - |C|^2
\rm .
\label{EL}
\end{equation}
Using Eqs.\,\eqref{Edelta}-\eqref{EC}, we find
\begin{equation}
C = t_1 t_2 \tau \sum_{j=0}^\infty p^j \exp\left[ -\frac{1}{2} (j \delta {- \eta})^2 \right]
\rm .
\label{ECexp}
\end{equation}
with the normalized offset variable $\eta=h/w_0$. It is normalized in the same way as $\delta$, and is a linear function of $\delta$ in the limits discussed here.

As it is obvious from the above, we use a very simple derivation for the walkoff loss here. Most importantly, in the definition \eqref{Epsi} of $\psi$ entering the overlap integral\,\eqref{EC}, we have ignored the effects of wavefront curvature and beam expansion of a more realistic gaussian beam. This is strictly only justified in the limits of very large beams and few reflections (small reflectivity) inside the etalon.

\begin{figure}[]
\centering\includegraphics[width=\columnwidth]{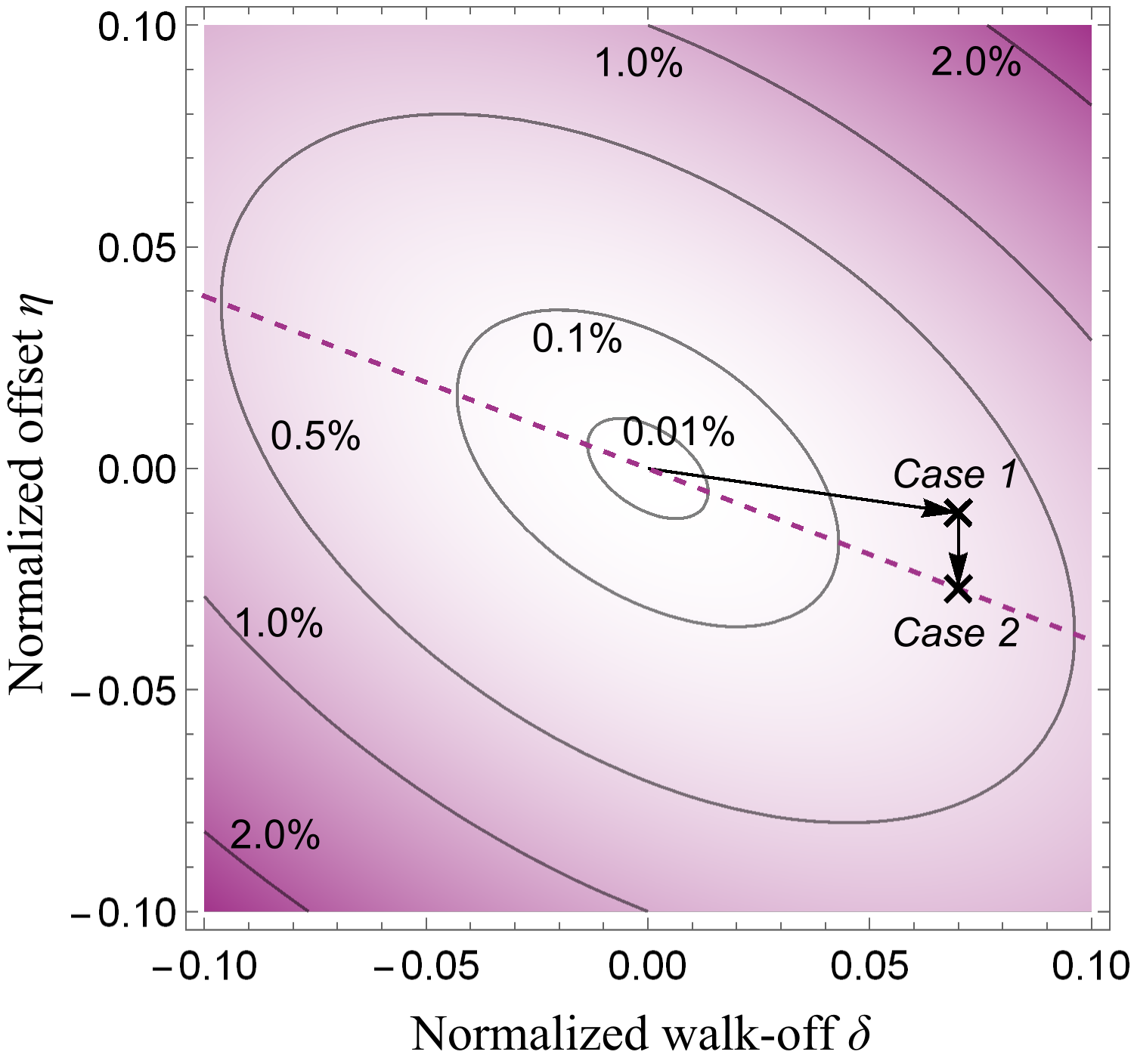}
\caption{(Color online) $L_{\rm{ gen}}(\delta,\eta)$ from Eq.\,\eqref{ELappr} for $R = 27\%$. (The numbers attached to the solid lines denote the magnitude of the walkoff loss in percent.) Eq.\,\eqref{ELappr} describes ellipses of equal loss in  $\lbrace\delta,\eta \rbrace$ space. When an etalon is inserted and tilted to a normalized angle $\delta$, it shifts the output beam to a normalized lateral offset $\eta(\delta)$ (\textit{Case 1}). When the laser cavity is realigned, the loss can be minimized, yielding $L_{\rm opt}(\delta)$ (\textit{Case 2}). By chosing the optimum reflectivity $R_{\rm opt}(n)$, the same optimum value can be obtained without realignment, yielding an effective "self-alignment" for all $\delta$  (dashed line).}
\label{FLVsetaVsdelta}
\end{figure}

We set $\tau = 1$, $r_1 = r_2 = \sqrt{R}$, and $t_1 = t_2 = \sqrt{T} = \sqrt{1-R}$ and find for the generic etalon loss in the limit of $\lbrace \delta, \eta \rbrace \ll 1$ 
\begin{equation}
L_{\rm{ gen}}(\delta, \eta) \simeq \eta^2 + \frac{2 R}{1-R} \eta \delta + \frac{R(1+R)}{(1-R)^2}\,\delta^2
\rm .
\label{ELappr}
\end{equation}
This expression differs from the expression found for the double-pass case\,\cite{Leeb1975,Korolev1976}, and the loss is in general larger than $1/2$ times the double-pass loss. Most notably, it contains the normalized lateral offset variable $\eta$, because unlike in the double-pass case, there is no mechanism of recycling parts of the higher-order reflection orders into the fundmantal mode\,\cite{Leeb1975,Korolev1976}. Eq.\,\eqref{ELappr} describes ellipses of equal loss in  $\left\lbrace\delta,\eta\right\rbrace$ space, see Fig.\,\ref{FLVsetaVsdelta}. It therefore needs in-depth examination, as we will present in the following Section.

\subsection{Case 1: Simple insertion loss}

In the most straight forward case, one will insert the etalon into the laser cavity under a given angle. This situation is indicated as the upper black arrow in Fig.\,\ref{FLVsetaVsdelta} (\textit{Case 1}), and we find the condition 

\begin{equation}
\eta_{\rm{ sim}}(\delta) \simeq -\frac{n-1}{2}\, \delta
\rm 
\label{ECon_del}
\end{equation}
for the normalized lateral offset. Under this condition, Eq.\,\eqref{ELappr} yields
\begin{equation}
L_{\rm{ sim}}(\delta) \simeq \left[\left(\frac{n-1}{2}\right)^2 - \frac{R(n-1)}{1-R} + \frac{R(R+1)}{(1-R)^2}\right]\,\delta^2
\rm .
\label{ELn}
\end{equation}
Such, we find a pure $\delta^2$ dependence known from the double-pass case\,\cite{Leeb1975,Korolev1976}. However, the design parameter dependend prefactor remains larger or equal to $1/2$ times the double-pass loss. Furthermore, we note that the loss goes with the square of the tilt angle $\theta$, or linearly with the peak transmission frequency of the etalon.

\subsection{Case 2: Insertion loss with optimum alignment}

If the laser cavity is realigned for maximum output power after etalon insertion, the loss can be reduced. In the course of the framework presented here, this is accomplished by minimizing \eqref{ELappr} with respect to the normalized lateral offset $\eta$ at constant normalized insertion angle $\delta$. In Fig.\,\ref{FLVsetaVsdelta} this corresponds to moving the point of operation from the upper black arrow's tip (\textit{Case 1}) to a lower value of $\eta$ (lower black arrow (\textit{Case 2}) in Fig.\,\ref{FLVsetaVsdelta})
\begin{equation}
\eta_ {\rm opt}(\delta) = \frac{h_ {\rm opt}}{w_0} \simeq {-} \frac{R}{1-R}\,\delta
\rm .
\label{Eetaopt}
\end{equation}
Eq.\,\eqref{Eetaopt} defines a line in $\lbrace\delta,\eta\rbrace$ space (dashed red line in Fig.\,\ref{FLVsetaVsdelta}). The minimized loss reads
\begin{equation}
L_ {\rm opt} \simeq \frac{R}{(1-R)^2}\,\delta^2 
\rm .
\label{ELopt}
\end{equation}
Thus, after alignment for loss minimization, the single pass loss equals exactly one half of the double pass loss\,\cite{Leeb1975,Korolev1976}. The double-pass case shows no dependence on alignment, and part of the higher reflection orders are recycled when passing the etalon for the second time.

\subsection{Loss minimization with refractive index}

As we have seen in the former Section, the insertion loss can be minimized by cavity realignment. The etalon's refractive index acts on the lateral walkoff through Snellius' law of refraction. Such, Eq.\,\eqref{ELn} suggests to trade $R$ against $n$ for loss minimization. This effective "self-alignment" condition is fulfilled when
\begin{equation}
R_{\rm opt} = \frac{n-1}{n+1}
\rm .
\label{ERopt}
\end{equation}
For this reflectivity, the optimum alignment condition \eqref{Eetaopt} is met, and such we find the minimized loss $L_{\rm opt}$. $R_{\rm opt}$ has no dependence on $\delta$, so the etalon fulfilling the design condition \eqref{ERopt} automatically provides the optimum alignment $\eta_{\rm opt}$ for all insertion angles, sparing the need for further manual alignment. Such, even when tuning the laser by tilting the etalon, the output power will remain at its optimum value. 

\begin{figure}[]
\centering\includegraphics[width=\columnwidth]{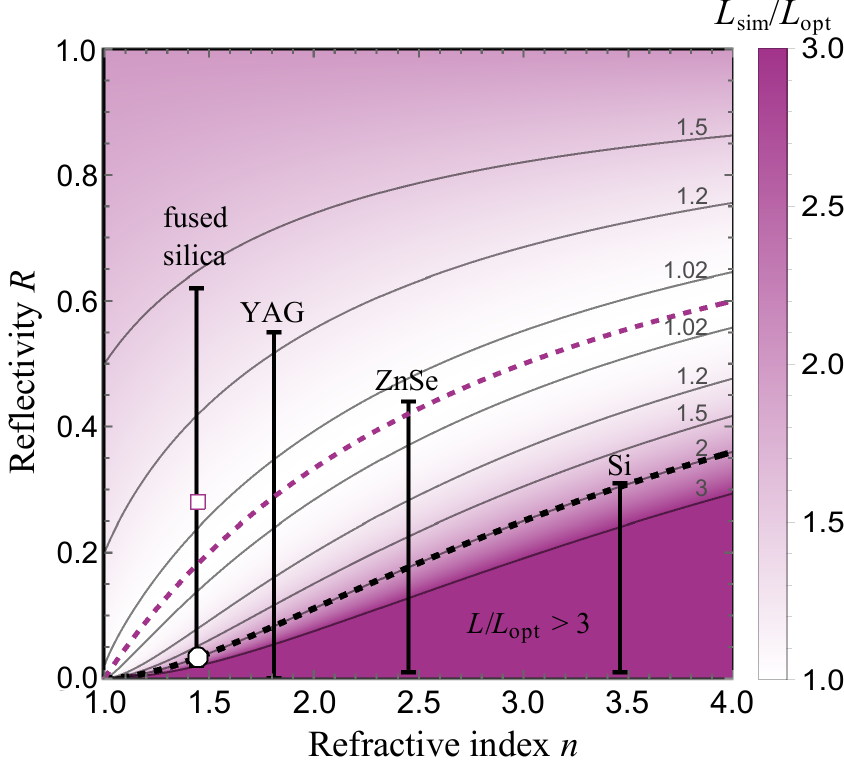}
\caption{(Color online) Ratio of the simple insertion loss $L_{\rm sim}$(Eq.\,\eqref{ELn}) over the optimized loss $L_{\rm opt}$(Eq.\eqref{ELopt}) as a function of $n$ and $R$. 
The purple dashed line is the optimized case with $R_{\rm opt}(n)$ given by Eq.\,\eqref{ERopt}.
The black dashed line is the result for Fresnel reflection $R_{\rm Fresnel}(n)$, Eq.\,\eqref{ERFresnel}.
The purple region in the bottom right corner of the plot stands for values $L_{\rm sim}/L_{\rm opt} > 3$. As discussed in Section\,\ref{SPracticalDesignConsiderations}, the solid black vertical bars indicate the refractive indices and reflectivity ranges which can be accessed using standard etalon materials and single-layer thin-film coatings (all values given from refractive indices at the common 1.55\,$\mu$m telecom wavelength).
The symbols (open circle = first etalon, open square = second etalon) refer to the laser design discussed in Section\,\ref{SExperiments}.
}
\label{FLVsRvsn}
\end{figure}

In Fig.\,\ref{FLVsRvsn}, we plot the ratio of the simple insertion loss (Eq.\,\eqref{ELn}) over the optimized loss (Eq.\eqref{ELopt}) as a function of $R$ and $n$. The dashed purple line indicates $R_{\rm opt}(n)$ from Eq.\,\eqref{ERopt}. We find that $R_{\rm opt}(n)$ is higher than the Fresnel reflection $R_{\rm Fresnel}(n)$ of the air-to-etalon transition, which is given by 
\begin{equation}
R_{\rm Fresnel}(n) \approx \left(\frac{n-1}{n+1}\right)^2 = R_{\rm opt}^2(n)
\rm ,
\label{ERFresnel}
\end{equation}
(dashed black line in Fig.\,\ref{FLVsRvsn}). The Fresnel reflectivity is the square of the optimum reflectivity, and we find $L_{\rm Fresnel} = 2 L_{\rm opt}$. Such, for all etalons a reflective coating is required to meet the "self-alignment" condition \eqref{ERopt}. It can be seen that without realignment, an etalon consisting of a dielectric medium will potentially introduce less walk-off loss than an air-spaced etalon with $n=1$, given that any etalon needs to be inserted under a minimum angle $\theta_{\rm min} > 0$.

\section{Numeric calculation}
\label{SNumericCalculation}

In order to get deeper insight into the crossover region between the analytic limits, we perform a numeric calculation. Following the ideas presented in~\cite{Leeb1975}, we calculate the overlap integral \eqref{EC} and minimize the loss by varying the value of the lateral offset $\eta$.

\begin{figure}[]
\centering\includegraphics[width=.95\columnwidth]{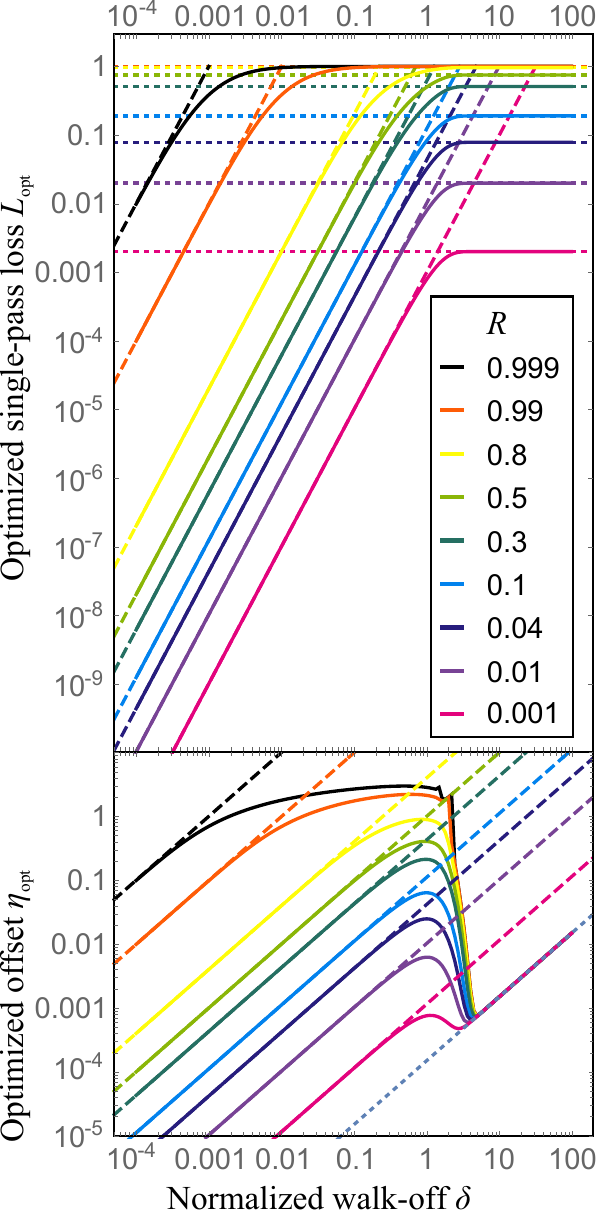}
\caption{(Color online)
(Upper panel) Calculation of the optimized single-pass walk-off loss, plotted as a function of the normalized beam displacement parameter $\delta$ for different values of the reflectivity ${R}$ as indicated in the legend. The corresponding analytic limits $L_{\rm max}$, $L_{\rm opt}$ from Eqs.\,\eqref{ELmax} and \eqref{ELopt} are plotted as dotted and dashed lines, respectively.
(Lower panel) The calculation delivers the values of the optimized offset $\eta_{\rm opt}$. The dashed lines are the analytical result for $\eta_{\rm opt}$ from Eq.\,\eqref{Eetaopt}. The dotted blue line is the simple insertion result $\eta_{\rm sim}$ from Eq.\,\eqref{ECon_del} for a refractive index of 1.0003, see text (legend valid for both panels).
}
\label{FEtalon_walk-off_loss}
\end{figure}

The numeric results are presented (parts presented in \cite{Eismann2012a}) for different surface reflectivities (set of solid lines) in Fig.\,\ref{FEtalon_walk-off_loss}. For $\delta \ll 1$, the loss rises quadratically with $\delta$, and the scaling factor given by the etalon surface reflectivity ${R}$, see Eq.\,\eqref{ELopt}. For comparison, we also plot the analytic limit for $L_{\rm opt}$ (dashed lines in Fig.\,\ref{FEtalon_walk-off_loss}) and find excellent agreement. 

Around $\delta \approx 1$ or $L_{\rm opt} \lessapprox 1$, whatever comes first, the analytic curves level off to the maximum value $L_{\rm max}(R)$ obtained analytically for complete separation of subsequent transmission orders in Eq.\,\eqref{ELmax} (dotted set of lines in Fig.\,\ref{FEtalon_walk-off_loss}). As mentioned before, the use of an intracavity etalon as a frequency filter becomes questionable in this regime.

Our numeric method yields the values of the optimized offset $\eta_{\rm opt}(R,\delta)$, which we show in the lower part of Fig.\,\ref{FEtalon_walk-off_loss}.
For small values of $R$ and $\delta$,  $h_{\rm opt} \propto \delta$. The dashed lines correspond to the value from the analytic limit\,\eqref{Eetaopt} and are in excellent agreement with the numerical optimization. 
At small values of $\delta$, it takes values $\eta_{\rm opt} \lessapprox 1$ for large values of $R$ only. This is intuitively clear, because for large reflectivities many orders of the summation in Eq.\,\eqref{ECexp} contribute to the coupling integral, and therefore shift the center of gravity of the wavefunction $\psi$.
In the crossover region defined by $L_{\rm opt}(\delta, R) \lessapprox L_{\rm max}(R)$, $\eta_{\rm opt}(\delta, R)$ reaches its maximum absolute value (note that in the coordinate system we introduced here, $\eta$ has negative values for positive $\delta$). For the largest values of $R \lessapprox 1$, the wavefunction $\psi$ has a flat-top shape. Also $\eta_{\rm opt} \approx 1$ and reaches a plateau, and the numeric optimization routine becomes slightly unstable, resulting in kinks of the curves in Fig.\,\ref{FEtalon_walk-off_loss}. The resulting $L_{\rm opt}(\delta, R)$ is almost constant, and very close to $L_{\rm max}(R)$. For $\delta > 1$, $\eta_{\rm opt}$ tends to the value of $\eta_{\rm sim}$, a behavior which is shown in the lower part of Fig.\,\ref{FEtalon_walk-off_loss} for a refractive index $n = 1.0003$\footnote{While approximating the value of the refractive index of air, this value has only been chosen to show the general tendency without dominating the offset $\eta_{\rm opt}$ for small values of $R$.}. This situation displays a large walk-off, and the optimization algorithm will re-align $\eta_{\rm opt}$ to the zero-order mode. As discussed before, the use of an etalon as a resonant frequency filter is questionable in this regime. 

In conclusion, for the numeric calculation presented here we find excellent agreement with the analytic cases in the limits of very small and very large walkoff. The agreement is found for both the optimized walkoff $\eta_{\rm opt}$, and the loss $L_{\rm opt}$. Furthermore, we gain additional insight in the crossover region where $L_{\rm opt}(\delta, R) \lessapprox L_{\rm max}(R)$.

\section{Practical design considerations}
\label{SPracticalDesignConsiderations}

We will treat here the typical use case of an etalon acting as an intra-cavity frequency filter for all but the desired (etalon- an laser-cavity-) resonant mode. The etalon is inserted in order to suppress the next-neighbor laser cavity mode from oscillating, the frequency distance between the two modes being ${\rm FSR}_{\rm las}$ (laser cavity free spectral range). The etalon therefore imposes a selection loss for the neighboring mode of

\begin{equation}
L_{\rm sel} \approx \frac{R}{(1-R)^2}\left( 2\pi\frac{{\rm FSR}_{\rm las}}{{\rm FSR}_{\rm eta}} \right)^2
\rm ,
\label{ELsel}
\end{equation}
with the etalon free spectral range ${\rm FSR}_{\rm eta} = c/2nd$ and $c$ the speed of light. Naively one would choose large values for the etalon reflectivity and its length, and use a highest-possible ${\rm FSR}_{\rm las}$ design. However, in practice one desires smallest possible walkoff loss for the oscillating mode, and largest possible suppression of the next-neighbor mode, which is given by the ratio

\begin{equation}
\frac{L_{\rm sel}}{L_{\rm opt}} \approx \left( \frac{2\pi^2 n^2 {\rm FSR}_{\rm las} z_0}{c} \right)^2
{\rm .}
\label{ELRatio}
\end{equation}
The ratio is independent of both $R$ and $d$, because both loss mechanisms show the same scaling in these two variables. Such, both design parameters can be chosen at will for providing the necessary side mode suppression. Therefore, this optimization does not hinder the possibility of picking a "self-alignment" combination of reflectivity and refractive index mentioned in Section\,\ref{SAnalytic_limits}. 
The remaining design parameters stem from the laser cavity design only: A larger laser cavity free spectral range ${\rm FSR}_{\rm las}$ requires a shorter laser cavity, and a large mode size at the etalon requires a large Raleigh range $z_0$.

The simplest etalon design has been discussed in Section\,\ref{SAnalytic_limits}, and consists of a blank substrate with a surface reflectivity given by the Fresnel reflectivity (Eq.\,\eqref{ERFresnel}). As it was shown, the ratio of the simple-insertion-loss-to-optimized-walk-off-loss is exactly 2, and leaves room for improvement.

A simple low-loss and cost-efficient design consists of a single-layer thin film coating on standard etalon substrates. If the coating's refractive index is higher than that of the substrate, the reflectivity is higher than in the uncoated case, and vice versa. A quarter-wavelength layer\footnote{A quarter wavelength layer has a physical thickness of exactly one quarter of the  wavelength of light travelling inside the layer. Quarter wavelength coatings are commonly used as anti-reflective coatings. However, the spectral width of the design is rather narrow, and the quarter-wavelength condition is strictly fulfilled for a given constant wavelength only.} provides highest (or lowest) reflectivity for a given combination of materials.

In Table\,\ref{T}, we give an overview on a selection of common etalon and coating material systems, and the resulting reflectivities and loss ratios $L_{\rm sim}/L_{\rm opt}$ (from Eqs.\,\eqref{ELn} and \eqref{ELopt}), based on refractive index data at 1.55\,$\mu$m\, from\cite{Polyanskiy2019}.  
It is interesting to note that by combining quarter-wave layers of commonly used oxides with fused silica (FS) or yttrium aluminum garnet (YAG) substrates, the loss ratio is very close to one, and the "self-alignment" condition Eq.\eqref{ERopt} is almost exactly met (combinations of FS + Ta$_2$O$_5$, FS + Nb$_2$O$_5$, YAG + TiO$_2$). A quarter-wavelength layer of amorphous silicon ($\alpha$-Si) on zinc selenide (ZnSe) delivers a similar result.
We did not consider the AR-coating material magnesium fluoride (MgF$_2$), which is given for completeness only.

Deviating from the quarter-wavelength condition allows one to continuously design any reflectivity between the extreme (quarter-wavelength) and the Fresnel reflectivity of the uncoated substrate by simply changing the layer thickness or the operation wavelength. In Fig.\,\ref{FLVsRvsn}, we plot the accessible reflectivity range for the materials shown in Table\,\ref{T} 
as vertical black solid bars. For all materials but silicon (Si), the optimum reflectivity $R_{\rm opt}$ (Eq. \eqref{ERopt}) can be reached by our design. Furthermore, we note that with more complicated coatings, almost any reflectivity can be produced. In the case of Si ($n=3.46$) as etalon substrate, a multi-layer coating would need to be designed in order to reach {$R_{\rm opt} = 0.55$}.

Finally, other etalon material properties such as bulk loss, surface loss, surface parallelism, quality of the coating and cost should be taken into account for any practical design. 

\begin{table*}[t]
\begin{center}
\begin{tabular}{ c  c | cc | cc | cc | cc | }

\hline\hline
&& \multicolumn{8}{c}{substrate material} \\
 &  & \multicolumn{2}{c}{fused silica} & \multicolumn{2}{c}{YAG} & \multicolumn{2}{c}{ZnSe} & \multicolumn{2}{c}{Si} \\ \hline
&	$n$  &  \multicolumn{2}{c}{1.44} & \multicolumn{2}{c}{1.81}& \multicolumn{2}{c}{2.45}& \multicolumn{2}{c}{3.46} \\ 
&	$R_{\rm uncoat}$     & \multicolumn{2}{c}{0.03}& \multicolumn{2}{c}{0.08}& \multicolumn{2}{c}{0.18}& \multicolumn{2}{c}{0.30} \\ \hline\hline
	
coating 	            & $n_{\rm coating}$ & $R_{\lambda/4}$ &$L_{\rm sim}/L_{\rm opt}$& $R_{\lambda/4}$ &$L_{\rm sim}/L_{\rm opt}$ & $R_{\lambda/4}$ &$L_{\rm sim}/L_{\rm opt}$ & $R_{\lambda/4}$&$L_{\rm sim}/L_{\rm opt}$\\ \hline
	MgF$_2$     & 1.37 & 0.02&3.28 & 0.00&498 & 0.02&28.5 & 0.09&13.2 \\
	silica & 1.46 & 0.04&1.81 & 0.01&24.5 & 0.01&107 & 0.06 & 22.6 \\ 
	Ta$_2$O$_5$ & 2.06 & 0.24&1.02 & 0.16&1.20 & 0.07&6.03 & 0.01&142 \\
	Nb$_2$O$_5$ & 2.24 & 0.31&1.08 & 0.22&1.04 & 0.12&3.30 & 0.03&40.5 \\ 
	TiO$_2$     & 2.43 & 0.37&1.14 & 0.28 & 1.00 & 0.17 &2.08 & 0.07 & 18.1 \\ 
	$\alpha$-Si           & 3.5 & 0.62&1.47 & 0.55&1.25 & 0.44&1.00 & 0.31&1.90 \\ 
	\hline\hline

\end{tabular}
\end{center}
\caption{Overview of combinations of four standard etalon substrate and six thin film coating materials (YAG = yttrium aluminum garnet). Refractive indices are given for the common 1.55\,$\mu$m telecom wavelength, see\,\cite{Polyanskiy2019} and references therein. The values for the coating's refractive indices $n_{\rm coating}$ are of indicative nature, since these may vary depending on the coating method and on the substrate. Quarter-wave layer reflectivities $R_{\lambda/4}$ follow from the refractive index combinations. The loss ratios $L_{\rm sim}/L_{\rm opt}$ follow from Eqs.\,\eqref{ELn} and \eqref{ELopt}.}
\label{T}
\end{table*}

\section{Experiments}
\label{SExperiments}

\begin{figure}[]
\centering\includegraphics[width=\columnwidth]{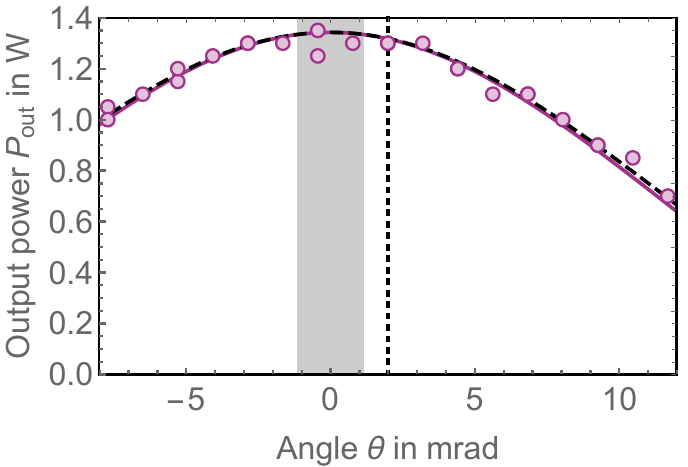}
\caption{(Color online) Tilt-tuning experiment with a laser setup as described in \cite{Eismann2012a}. For different values of the tilt angle $\theta$, the output power changes due to a change in walk-off loss from the etalon. The solid line is a constrained fit of the theory with simple insertion walk-off loss, and the dashed line indicates the possible improvement by realignment or by choice of the "self-alignment" condition from Eq.\,\eqref{ELopt}. The shaded area indicates tilt angles smaller or equal to the minimum insertion angle $|\theta| \leq \theta_{\rm min}$, and the dotted line indicates the chosen point of laser operation at $\theta = 2.0$\,mrad.
}
\label{FExp}
\end{figure}

In order to test our theory results, we perform an etalon tilt-tuning experiment inside an active laser cavity. We measure the output power of a diode-pumped all-solid-state laser emitting at $\lambda = 1342$\,nm, see \cite{Eismann2012} for details of the setup. Single frequency operation of the unidirectional ring cavity is enforced by inserting two etalons. 

The first etalon is used to select an operation region inside the laser's gain profile, which has a bandwidth of $\approx$\,300\,GHz. The etalon is made of uncoated IR fused silica, $n = 1.447$ at 1342\,nm\,\cite{Leviton2007}, and $d = 0.5$\,mm yielding a free spectral range of 210\,GHz. Relying on Fresnel reflection, we have $R = 3.3\%$, and found a loss ratio of $L_{\rm sim}/L_{\rm opt} = 2$ in Section\,\ref{SNumericCalculation}. However, when tilt-tuning the first etalon by one of its free spectral ranges, the imparted walk-off loss $L_{\rm sim}$ is below one per mille, and can therefore be neglected compared to the total round-trip loss. The smallness of the loss is verified by tilt-tuning the first etalon over multiples of its free spectral ranges at constant operation frequency without finding a significant reduction in output power.

The second etalon is used to enforce single-mode operation, and for fine-tuning of the laser inside the operating region. It has a length of $d = 4$\,mm, corresponding to a free spectral range of ${\rm FSR}_{\rm eta} = 26$\,GHz. 
The etalon is made of infrared fused silica, coated with a single quarter-wavelength layer of tantalum pentoxide (Ta$_2$O$_5$). This simple low-loss coating results in a reflectivity of $R = 27\%$ at 1342\,nm. 
With $L_{\rm sim}/L_{\rm opt} \approx 1.05$ the design is very close to the "self-alignment" condition, see Section\,\ref{SAnalytic_limits}, and therefore well suited for tilt-tuning. It is mounted in a temperature-stabilized copper enclosure which can be tilt-tuned using two fine-thread screws around the horizontal and the vertical axis, and a piezoelectric transducer (PZT) around the vertical axis, see \cite{Eismann2012a} for details. 

The variations in the gain spectrum\,\cite{Eismann2013} and the residual loss modulation by the first etalon are negligible for the small values of tuning of less than 8\,GHz we apply here. We find a selection loss ratio $L_{\rm sel, 2}/L_{\rm sel, 1} \approx 930$ between second and first etalon, see Eq.\,\eqref{ELsel}. Such, in what follows we will safely neglect all spectral dependences of the round-trip loss (and gain) apart from the contribution of the second etalon only. Both etalon designs are shown as open symbols in Fig.\,\ref{FLVsRvsn}.

In order to find a suitable point of operation for the second etalon, it is tilted around both axes for setting the etalon perpendicular to the incident beam. Here, $\theta < \theta_{\rm min} \approx 1.2$\,mrad, and we find the unstable operation characteristics mentioned in Section\,\ref{SIntroduction} ({shaded area in Fig.\,\ref{FExp}}). For the determination of the minimum insertion angle $\theta_{\rm min}$, we have used the beam waist value $w_0 = 370\,\mu$m, which can be calculated from backpropagation of the measured laser output beam profile to the intracavity position of the etalon\,\cite{Eismann2012a}. 

We calibrate the tuning angle as a function of vertical screw turns by measuring the emission frequency of the laser, using frequency-doubled light and a \textit{High Finesse WS-6} wavelength meter, as described in Ref.\,\cite{Eismann2012}. 
The calibration reveals a value of $(9.7 \pm 0.1_{\rm stat})$\,mrad per turn, which is in good agreement with the value obtained from the geometry of the mount, and the screw's thread dimensions. During normal operation, the etalon position is sufficiently larger than $\theta_{\rm min}$ in order to stay well within the regime of stable operation, but with negligible power loss. We use an offset angle of $\theta_0 \approx 2.0\,{\rm mrad}~\widehat{\approx}~1.7\,\theta_{\rm min}$, a value allowing for longterm stable operation of the laser.

The output power of the laser is given by \cite{Eismann2012,Rigrod1963}
\begin{equation}
P_{\rm out}(\theta) = P_{\rm sat}T_{\rm out}\left[\frac{G_0}{T_{\rm out}+L_{\rm tot}(\theta)}-1\right]
\label{EP}
\end{equation}
with the saturation power $P_{\rm sat}= 44$\,W, the small-signal gain $G_0 = 0.11$, and the output coupler transmission $T_{\rm out}=3.5\%$, all of which are known from the laser design, see Ref.\,\cite{Eismann2012}. The total intracavity loss $L_{\rm tot}(\theta) = L_0 + L(\theta)$ consists of a constant term $L_0$ (taking into account all residual thermal lens loss, scattering, bulk absorption and imperfections of AR coatings of the various intracavity elements) and a tilt-angle dependent term $L(\theta)$ which we intend to measure.  Here we use the definition of the simple insertion loss $L(\theta) = L_{\rm sim}(\theta)\approx 1.264\times 10^{-4}$\,mrad$^{-2} \times \theta^2 $ given by Eq.\,\eqref{ELn} (the laser cavity is not realigned after tuning), with the design data from the second etalon and $w_0 = 370\,\mu$m as above.

We plot the experimental data in Fig.\,\ref{FExp}. 
The purple solid line is a fit of the data to Eq.\,\eqref{EP} with $L_0$ as the only free fit parameter. The maximum output power of $1.3$\,W is determined by the cavity roundtrip loss $L_0 = (2.13 \pm 0.03_{\rm stat})\%$. This offset loss in the few-percent range is expected in our design, which uses several intra-cavity elements with their respective insertion loss. $L_0$ is probably dominated by the contribution of the thermal lens loss in the active medium, see for instance Ref.\,\cite{Eismann2013} and references therein.

The inverted-parabola-like shape of the output power as a function of $\theta$ is entirely given by the theory presented here. The data and theory are in remarkable agreement. A fit of the data with $P_{\rm sat}$, $G_0$ as additional free fit parameters yields similar values for all parameters, however with significantly higher statistical error bars.

The laser cavity was not realigned after tilting the etalon from the offset position, therefore one can gain output power by realignment, since the "self-alignment" condition described in Section\,\ref{SAnalytic_limits} is almost but not quite met ($L_{\rm sim}/L_{\rm opt}\approx 1.05$ for the second etalon). 
The relevant output power curve with $L_{\rm opt}(\theta)/\theta^2 \approx 1.203\times 10^{-4}$\,mrad$^{-2}$ given by Eq.\,\eqref{ELopt} is plotted as a dashed black line in Fig.\,\ref{FExp}. As a figure of merit, the possible output power improvement given by realignment is 0.1\% at the operation point $\theta_0$. Since this improvement is within the error bar of the power measurement, we consider the realignment unnecessary for this mode of operation. However, when tuning to the maximum tilt angle of 12\,mrad, a 4\% power increase can potentially be achieved.

In order to achieve highest possible output powers at the given fixed frequency of choice, the etalon is tilted to $\theta_0$, giving rise to $L_{\rm sim} \approx 0.05\%$. \footnote{If this contribution to loss could be avoided, approx. 2\% of output power could be gained.} The laser frequency is then adjusted by temperature tuning of the etalon, as described in Ref.\,\cite{Eismann2012a}. We note that more than 10\,GHz of temperature tuning across the same frequency range were demonstrated, with the laser output power remaining constant within the error bar of the measurement. This observation gives confidence in the assumptions of a flat gain spectrum and an almost flat response of the first etalon along the tuning range, which were taken above. 

Mode-hop free tuning of the laser is achieved by changing the laser cavity length with a linear voltage ramp applied to a PZT holding one of the cavity mirrors. Since the scan frequency is chosen to be 1\,Hz or faster, applying a temperature ramp to the second etalon is not an option. Therefore, the same ramp voltage is fed to the tilt PZT of the second etalon, with a suitable linear feed-forward factor. Such, mode-hop free tuning of up to 1.1\,GHz could be achieved, however at the price of a $\approx 10\%$ power drop of the laser at the highest tilt angle.

We note that in this practical laser design, the loss introduced by etalon tilt-tuning can become the dominant intracavity loss term, strongly reducing output power to approximately one half of the maximum power. Furthermore, the striving agreement we find between theory and the experiment in this Section strengthens our confidence in the analytic theory in presence of the strong simplifications which were applied.

\section{Conclusion}
\label{SConclusion}

In this article, we have presented a simple analytic model describing Fabry-P\'erot etalon walk-off loss for optimizing the design of lasers and photonic devices. We found a quadratic scaling of the simple insertion loss $L_{\rm sim}$ with the normalized walk-off $\delta$. Furthermore, we found that the loss can be minimized either by realignment of the laser cavity, or by chosing a combination of etalon refractive index and surface reflectivity in such a way that the realignment happens automatically, giving $L_{\rm opt} \leq L_{\rm sim}$. We performed a numeric study of the loss beyond the approximations of small walkoff taken for the analytic model, and find excellent agreement between analytics and numerics. We discussed important implications on practical lasers, and found that loss-optimized etalons can be realized using very simple designs of standard materials and single-layer thin film coatings. Finally, we performed a tilt-tuning experiment inside an active laser cavity, and find excellent agreement between theory and the output power data.

This work should give rise to improved and more efficient laser designs, and in general to improved optical setups using etalons. A potential avenue for future investigation is the role of tuning in the etalon design process.

\section*{Acknowledgements}
Acknowledgements: The authors wish to acknowledge discussions with Jacques Vigu\'e, G\'erard Tr\'enec, Christophe Salomon, Andrea Bergschneider, Felix Rohde, Peter Zimmermann und Frank Wunderlich.

\bibliographystyle{unsrt}
\bibliography{Etalon_loss}

\begin{thebibliography}{10}

\bibitem{Peterson1966}
Don~G. Peterson and Amnon Yariv.
\newblock Interferometry and laser control with solid fabry-perot etalons.
\newblock {\em Appl. Opt.}, 5(6):985--991, Jun 1966.

\bibitem{Eismann2013}
U.~Eismann, A.~Bergschneider, F.~Sievers, N.~Kretzschmar, C.~Salomon, and
  F.~Chevy.
\newblock 2.1-watts intracavity-frequency-doubled all-solid-state light source
  at 671 nm for laser cooling of lithium.
\newblock {\em Opt. Express}, 21(7):9091--9102, Apr 2013.

\bibitem{Miake2015}
Yudai Miake, Takashi Mukaiyama, Kenneth~M. O'Hara, and Stephen Gensemer.
\newblock A self-injected, diode-pumped, solid-state ring laser for laser
  cooling of li atoms.
\newblock {\em Review of Scientific Instruments}, 86(4):043113, 2015.

\bibitem{Koch2015}
Peter Koch, Felix Ruebel, Juergen Bartschke, and Johannes~A. L'huillier.
\newblock 5.7 w cw single-frequency laser at 671 nm by single-pass second
  harmonic generation of a 17.2 w injection-locked 1342 nm nd:yvo$_4$ ring
  laser using periodically poled mgo:linbo$_3$.
\newblock {\em Applied optics}, 54(33):9954--9959, 2015.

\bibitem{Cotteverte1991}
JC~Cotteverte, Fabien Bretenaker, and A~Le Floch.
\newblock Jones matrices of a tilted plate for gaussian beams.
\newblock {\em Applied optics}, 30(3):305--311, 1991.

\bibitem{Abu-Safia1994}
H~Abu-Safia, R~Al-Tahtamouni, I~Abu-Aljarayesh, and NA~Yusuf.
\newblock Transmission of a gaussian beam through a fabry-perot interferometer.
\newblock {\em Applied optics}, 33(18):3805--3811, 1994.

\bibitem{Nichelatti1995}
Enrico Nichelatti and Gianemilio Salvetti.
\newblock Spatial and spectral response of a fabry--perot interferometer
  illuminated by a gaussian beam.
\newblock {\em Applied optics}, 34(22):4703--4712, 1995.

\bibitem{Danielmeyer1970}
H.~G. Danielmeyer.
\newblock Stabilized efficient single-frequency nd:yag laser.
\newblock {\em Quantum Electronics, IEEE Journal of}, 6(2):101--104, 1970.

\bibitem{Korolev1976}
F.A. Korolev, L.E. Grin', P.V. Korolenko, V.V. Lebedeva, A.I. Odintsov, and
  N.\'{E} Sarkarov.
\newblock Losses of a laser resonator with an inclined fabry-perot etalon as a
  frequency selector.
\newblock {\em Journal of Applied Spectroscopy}, 25(6):1506--1509, 1976.

\bibitem{Leeb1975}
Walter~R. Leeb.
\newblock Losses introduced by tilting intracavity etalons.
\newblock {\em Applied Physics A: Materials Science \& Processing}, 6:267--272,
  1975.
\newblock 10.1007/BF00883762.

\bibitem{Eismann2019}
U.~Eismann, W.C. Hurlbut, D.B. Foote, D.J. Christensen, M.J. Cich, and
  C.~Haimberger.
\newblock Broadband high-resolution spectroscopy in the mid infrared via a
  tunable, continuous-wave optical parametric oscillator.
\newblock In {\em Proceedings of ICOLS 2019, Queenstown, New Zealand}, 2019.

\bibitem{Murphy2007}
M.~T. Murphy, Th. Udem, R.~Holzwarth, A.~Sizmann, L.~Pasquini, C.~Araujo-Hauck,
  S.~Dekker, H.and~D'Odorico, M.~Fischer, T.~W. H\"ansch, and A.~Manescau.
\newblock High-precision wavelength calibration of astronomical spectrographs
  with laser frequency combs.
\newblock {\em Monthly Notices of the Royal Astronomical Society},
  380(2):839--847, 08 2007.

\bibitem{Tian2016}
Long Tian, Shujing Li, Haoxiang Yuan, and Hai Wang.
\newblock Generation of narrow-band polarization-entangled photon pairs at a
  rubidium d1 line.
\newblock {\em Journal of the Physical Society of Japan}, 85(12):124403, 2016.

\bibitem{Eismann2012a}
Ulrich Eismann.
\newblock {\em A novel all-solid-state laser source for lithium atoms and
  three-body recombination in the unitary Bose gas}.
\newblock PhD thesis, Universit\'e Pierre et Marie Curie -- Paris VI, 2012.

\bibitem{Polyanskiy2019}
Mikhail~N. Polyanskiy.
\newblock Refractive index database.
\newblock \url{https://refractiveindex.info}.
\newblock Accessed on 2019-08-23.

\bibitem{Eismann2012}
U.~Eismann, F.~Gerbier, C.~Canalias, A.~Zukauskas, G.~Tr\'enec, J.~Vigu\'e,
  F.~Chevy, and C.~Salomon.
\newblock An all-solid-state laser source at 671 nm for cold-atom experiments
  with lithium.
\newblock {\em Applied Physics B}, 106:25--36, 2012.
\newblock 10.1007/s00340-011-4693-y.

\bibitem{Leviton2007}
D.~B. Leviton, B.~J. Frey, and T.~J. Madison.
\newblock Temperature-dependent refractive index of caf2 and infrasil 301.
\newblock {\em Proc. SPIE 6692, 669204}, 2007.

\bibitem{Rigrod1963}
W.W. Rigrod.
\newblock Gain saturation and output power of optical masers.
\newblock {\em Journal of Applied Physics}, 34(9):2602--2609, 1963.

\end{thebibliography}

\end{document}